\documentclass[]{aa}
\usepackage{graphicx}
\usepackage{txfonts}
\usepackage{natbib}
\def\teff{$T\rm_{eff }$}
\def\kms{$\mathrm {km s}^{-1}$}
\begin{document}
   \title{Family ties: abundances in Terzan 7, a Sgr dSph globular cluster
\thanks{Based  on observations obtained in ESO programmes 67.B-0147 and 65.L-0481}
}

   \author{L. Sbordone \inst{1,2},
           P. Bonifacio \inst{3},
           G. Marconi \inst{1,4}
           \and
           R. Buonanno \inst{2}
	   \and
	   S. Zaggia \inst{3}
          }

   \offprints{L. Sbordone}

   \institute{ESO  European Southern Observatory --
              Alonso de Cordova 3107 Vitacura, Santiago, Chile
             \and
             Universit\'a di Roma 2 ''Tor Vergata'' --
             via della Ricerca Scientifica, Rome, Italy
             \and
             Istituto Nazionale di Astrofisica --
             Osservatorio Astronomico di Trieste, Via Tiepolo 11,
             I-34131 Trieste, Italy
             \and
             Istituto Nazionale di Astrofisica --
             Osservatorio Astronomico di Roma
             Via Frascati 33, 00040 Monte Porzio Catone, Rome, Italy
             }

   \date{Received ; accepted }

   \abstract{
We study the chemical composition of 5 giant stars in the globular
cluster \object{Terzan 7} associated with the Sagittarius dwarf spheroidal galaxy (\object{Sgr dSph}), based on high resolution UVES-VLT spectra.
We confirm the metallicity found by  previous high resolution investigations: [Fe/H]$\sim -0.6$.
We also show that this cluster displays the same low
$\alpha-$element to iron ratio found in \object{Sgr dSph} field stars of similar
metallicity, as well as the same low Ni/Fe ratio.
These chemical signatures are  characteristic of the \object{Sgr dSph} system, 
and appear to be shared both by the globular cluster 
\object{Pal 12}, which was most likely stripped from Sgr
by tidal interaction, and by \object{Pal 5}, which may also have belonged in the past to the \object{Sgr dSph} system. Intriguingly even globular cluster
\object{Ruprecht 106}, although not associated to Sgr, displays
similar characteristics.
   }
\authorrunning{L.Sbordone et al.}
\titlerunning{Abundances in Terzan 7}
   \maketitle

\section{Introduction}
\object{Terzan 7} (\citealt{terzan,buonanno95}) is a sparse, young, and metal
rich globular cluster,
physically associated to the Sagittarius dwarf spheroidal galaxy
(\object{Sgr dSph}, \citealt{ibata95,marconi98}). 
The dwarf galaxy, to which 4 globular clusters 
are dynamically linked
(\object{M54}, \object{Terzan 7}, \object{Terzan 8}, and \object{Arp 2}), is
undergoing tidal merging with the Milky
Way (MW) Halo, and is gradually being disrupted. Clear detection of the
associated stellar stream inside the Halo has been achieved with 2MASS data
(\citealt{majewski03}); clear indications exist that some MW globular clusters
most likely originated in the \object{Sgr dSph} system, and then were stripped and added to
the halo GC system (the best candidate being \object{Pal 12}, see \citealt{cohen03}).

\object{Terzan 7} has long been known to be a somewhat peculiar globular
cluster. Even before
the \object{Sgr dSph} discovery (\citealt{ibata95}), it has been associated with a small
group of other objects due to their anomalous characteristics. It was shown to
share with \object{Arp 2}, \object{Ruprecht 106} and \object{Pal 12} the young age and a significant
discrepancy between the photometric metallicities and those based on the 
\ion{Ca}{ii} IR
triplet (\citealt{buonanno95} and references therein). The fact that all these
clusters, with the exception of \object{Ru 106}, now appear to be somehow linked to the
Sgr dSph leads one to interpret their anomalies as linked to their origin in the \object{Sgr
dSph} system.

We are conducting a comprehensive study of chemical abundances in the \object{Sgr dSph}
system, in order to clarify its evolutionary characteristics, along with the
effect of its interaction with the MW Halo. In previous papers we presented a
study of 12 giant stars in the main body of \object{Sgr dSph}  (\citealt{bonifacio00},
\citealt{bonifacio03}, henceforth Paper I), while in this paper we present a similar
analysis of five \object{Terzan 7} giants.

As shown in Paper I, we identified a very young, metal rich
population ([Fe/H]$\sim$0, age less than 2 GYr) in the \object{Sgr dSph}. This population was also
characterized by relatively low $\alpha $ over iron ratios. Such feature may
be interpreted as the signature of a prolonged, slow, or bursting star formation.
A trend in [$\alpha$/Fe] vs. [Fe/H] towards lower $\alpha$ content at high iron
content was barely detectable in our sample, but emerged clearly when the
abundances of \object{M54} giants were also included (\citealt{brown99}). This was even more
evident by looking at all the available data of the Local Group dwarf galaxies.
This trend appears to be significantly different from the one displayed by the
MW stars, which leads us to infer a different evolutionary behavior between
Local Group dwarf spheroidals and the Milky Way. In this scenario, the
chemical composition 
of \object{Terzan 7} (and of the other globular clusters associated to the Sgr
dSph) is of particular importance; since it belongs to the \object{Sgr dSph} system, we
expect to find the same chemical peculiarities we observed in the main body of
the galaxy.

\begin{table*}
\begin{center}
 \caption{Photometry and physical parameters for the five stars.}
\begin{tabular}{llllllccc}
\hline
{\bf Number$^a$} & {\bf V} & {\bf (B-V)}{\boldmath $_0$} &
{\bf T}{\boldmath $_{eff}$} & {\boldmath $\log{g}$} & {\boldmath $\xi$} & $\alpha(J2000)$& $ \delta (J2000)$ & ${v_{r}}^b$\\
 & mag & mag & K & cgs & \kms & hms & $^\circ,',''$ & \kms \\
\hline
 \object{1272}     & 16.62 & 1.15 & 4421 & 1.2 & 1.45 &19 17 37.1 & -34 39 11.9 & 158.9 \\
 \object{1282}$^c$ & 16.08 & 1.30 & 4203 & 1.3 & 1.60 &19 17 39.4 & -34 39 06.4 & 158.2 \\
 \object{1515}     & 16.76 & 1.12 & 4468 & 2.0 & 1.45 &19 17 38.2 & -34 39 16.8 & 157.8 \\
 \object{1665}$^d$ & 15.04 & 1.50 & 3945 & 0.8 & 1.60 &19 17 43.2 & -34 39 43.2 & 159.6\\
 \object{1708}$^e$ & 16.08 & 1.28 & 4231 & 1.2 & 1.70 &19 17 43.5 & -34 39 12.4 & 160.6\\
\hline
\\
\multispan{6}{{\scriptsize $a$ Referred to the catalog of Buonanno et al. (1995), available via CDS} \hfill}\\
\multispan{6}{{\scriptsize ~~~at {\tt http://cdsweb.u-strasbg.fr/cgi-bin/qcat?J/AJ/109/663/}. }\hfill}\\
\multispan{6}{{\scriptsize $b$ The presented $v_r$ is the heliocentric value.} \hfill}\\
\multispan{6}{{\scriptsize $c$ This is star \object{S34} in \citet{tarantella04}} \hfill}\\
\multispan{6}{{\scriptsize $d$ This is star \object{S16} in \citet{tarantella04}} \hfill}\\
\multispan{6}{{\scriptsize $e$ This is star \object{S35} in \citet{tarantella04}} \hfill}\\
\\
\end{tabular}
\label{tabstars}
\end{center}
\end{table*}

   \begin{figure}
   \centering
   \includegraphics[width=9cm]{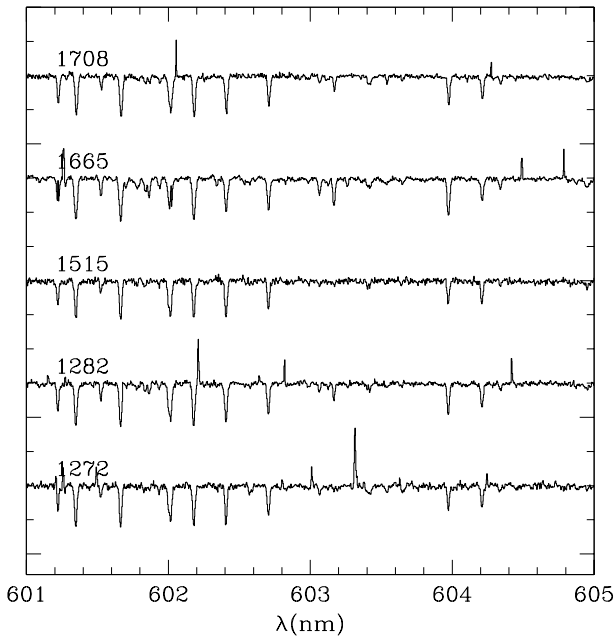}
      \caption{A sample of the spectra of the five stars in the wavelength range 601 to 605 nm, where many of the considered \ion{Fe}{i} features lie. For Star \# \object{1282} the DIC1 spectrum is shown.
              }
         \label{spettri}
   \end{figure}

Detailed abundances on \object{Terzan 7} stars are presented in the literature only by
\citet{tarantella04}, where the authors analyzed three cool  low-gravity stars,
one of which could be an AGB due to its very low gravity,
and found a rather high mean metallicity ([Fe/H]$\sim$-0.6), very slight
$\alpha$
enhancement, and a marginally significant Ni under-abundance with respect to
iron. A strong La and Eu overabundance is also noticeable in their results.

Along with the Sgr dSph main body stars described in \citet{bonifacio03}, we
also observed three giants in \object{Terzan 7} (see \citealt{sbordone04}).
One of the stars
of
\citet{tarantella04} was also
observed by us, although with a slightly different
instrumental setting.
We decided to produce a methodologically coherent analysis of
all the five \object{Terzan 7} stars
observed at high resolution,
looking in particular for the presence of a ``family
signature'' linking Terzan 7 with the Sgr dSph and the other globular clusters
associated to the system.

\section{Observations and data reduction}

The five stars studied in this paper have all been observed 
at the VLT/UVES high resolution spectrograph.
Their coordinates, photometry,  and atmospheric 
parameters are listed in Table
\ref{tabstars}.
The first three stars (\# \object{1272}, \# \object{1515}, \# \object{1282})  have been observed
in the course of our programme in dichroic 1 (DIC1)
mode during the same session devoted to the \object{Sgr dSph} main body described in
Paper I. Both blue and red arms were used, but only red arm data
is presented here. The integration time was 3600~s on each star, 
split into two 1800~s
exposures.The pipeline extracted, radial velocity corrected spectra were
coadded to reach a S/N$\simeq50$ at 600~nm, increasing at longer
wavelengths.

The spectra for the three stars described in \citet{tarantella04} were
retrieved from the
ESO archive. They are dichroic 2 spectra (DIC2), with both blue and red
arm data (blue chip 375 to 500~nm, red CCD mosaic
593 to 960~nm).
Each
star received a single 3600~s exposure. We have extracted the spectra by means
of the UVES pipeline, obtaining an S/N of about 70 at 600~nm.
Blue DIC 2 chip was not in fact used: some lines in that wavelength range were initially
included in the linelist, but in metal rich, low gravity stars that area of the spectrum is subject to such crowding that a
proper continuum estimate was deemed very difficult and
uncertain.

A sample of the spectra is shown in Fig. \ref{spettri}.
The star identified as S34 in \citet{tarantella04} is
Star \# \object{1282}, also observed independently by us.


For all the stars, equivalent widths (EWs) were measured for lines of \ion{Mg}{i},
\ion{Si}{i}, \ion{Ca}{i}, \ion{Ti}{i}, \ion{Fe}{i}, \ion{Fe}{ii} and \ion{Ni}{i} by using {\tt IRAF} task {\tt splot}; see
Table \ref{abund1}  for details. The lines employed are
essentially the same used in Paper I for \# \object{1515} and \# \object{1272}, while a completely
different linelist was prepared for stars \# \object{1665} and \# \object{1708},  due to the different
spectral coverage. Star \# \object{1282}, having both DIC1 and DIC2 spectra available, had
the two linelists merged. In the (sporadic) case of a line being measured in
both spectra, the mean of the two EWs was used.


\section{Abundance analysis}

The analysis method is the one already described
in Paper I: the effective temperatures have been derived from $(B-V)_0$ colors
using the calibration by \citet{alonso99}; we assumed E(B-V)=0.07 as in
\citet{marconi98}. Notice that the calibration provided in \citet{alonso99} for
$(B-V)_0$ colors is marginally dependent on the star's [Fe/H]. For
verification, we first derived T$_{eff}$ assuming [Fe/H]=-0.6, and then
recalculated by using the derived metallicity. The variations were
totally negligible, on the level of a few K.

\citet{tarantella04} used a different approach to T$_{eff}$
determination and relied on an
excitation temperature; nevertheless their estimates result in
good agreement with
our \teff 's , differing at most
50K. Estimates of the effect of systematics in the physical parameters
are given in Table \ref{errors}.

  \begin{figure}
  \centering
  \includegraphics[width=\hsize]{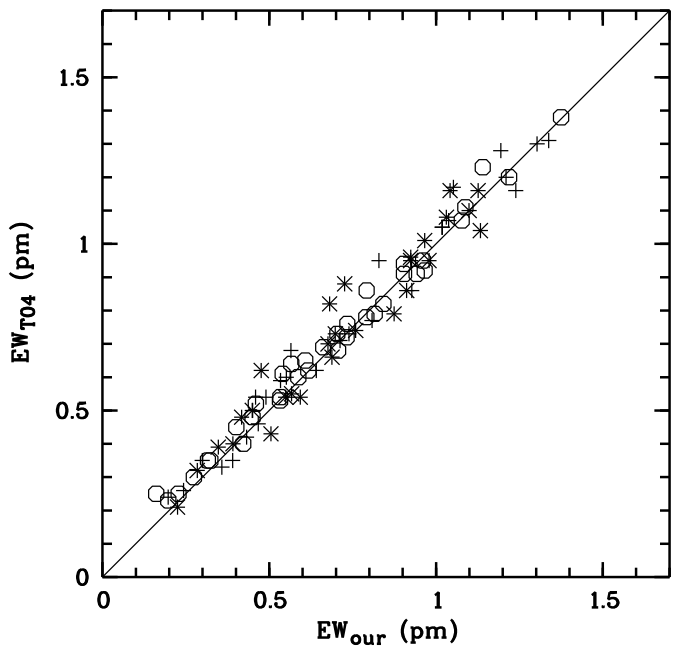}
     \caption{Comparison of the equivalent widths
measured by us and by \citet{tarantella04}
for the lines in common. Open hexagons  refer to Star \# \object{1282},  crosses refer to Star \# \object{1665}
and asterisks to Star \# \object{1708}. The straight line is not
a fit, but simply the bisector.}
        \label{pl_ew}
  \end{figure}

\begin{figure}
\centering
\resizebox{8cm}{!}{\includegraphics{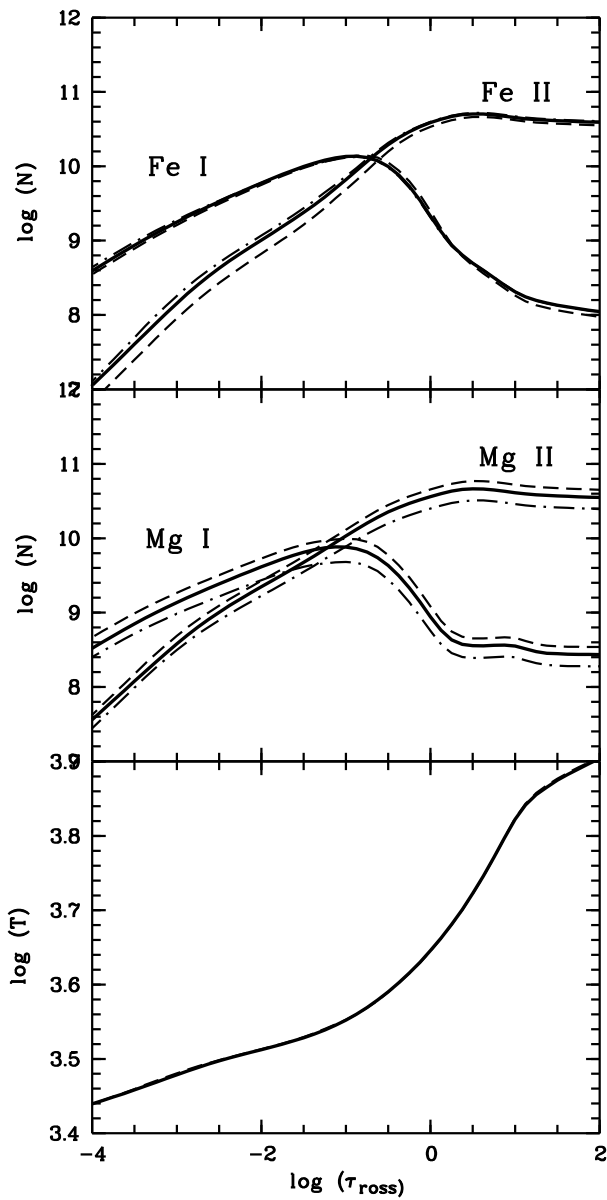}}
\caption{Atmospheric models for Star \# \object{1665}. All models correspond
to \teff = 3945 K log g = 0.80. The solid lines refer to an
opacity sampling ATLAS 12 model with abundances as provided in Table
\ref{abund_abs}. The dashed lines refer to an ATLAS 9
model computed with a solar-scaled ODF with [M/H]=-0.5 and
a microturbulence of 1 \kms. The dash-dotted lines refer to the
ionization structure of the ATLAS 9 model computed by WIDTH
when the abundances of Table \ref{abund_abs} are provided as input.
The top panel shows the \ion{Fe}{i} and \ion{Fe}{ii} ions, the middle panel
the \ion{Mg}{i} and \ion{Mg}{ii} ions, while the bottom panel displays the temperature
structure of the model. The neutral ions show a peak in number
densities around $\log\tau_{ross}\sim -1.5$. }
\label{models}
\end{figure}

Gravity was initially estimated from a comparison with Padova
isochrones (\citealt{salasnich00}). One-dimensional, LTE models were
calculated by using the ATLAS 9 code (\citealt{kurucz93}). The abundances were
extracted from the measured EWs by using WIDTH; spectral syntheses were produced with SYNTHE (for both codes see \citealt{kurucz93}).
SYNTHE was first used in the DIC 2 linelist selection phase when it helped in selecting
unblended lines of adequate strength, and then after the analysis, as a check of internal consistency
by verifying correspondence between the measured lines and their syntheses. For all these codes, our Linux ported versions were
used, as described in \citet{sbordone04_3}.

The models were calculated without introducing overshooting (see 
\citealt{castelli97}), and with solar scaled abundances, since no significant
$\alpha$-enhancement was found. Where necessary, the original estimated
gravity has been adjusted in order to satisfy the iron ionization equilibrium.
The changes were generally small, 
on the order of $\Delta \log g \sim 0.1$. 
A notable exception is Star \# \object{1272} which,
with magnitude and colours similar to 
Star \# \object{1515}, has a gravity coherent 
with stars 
that are almost 0.5
mags brighter in V and 0.15 mags redder in (B-V), like, for example, Star \#
\object{1282}.
For Star \# \object{1272} the spectroscopic gravity is 
therefore about 0.8 dex {\em lower} 
than the one suggested by the isochrones.

 The microturbulence was determined, as usual, by
requiring no correlation between the EW and the derived abundance for each \ion{Fe}{i}
line.

For one star,
\# \object{1665}, significant inconsistencies emerged between the abundances
computed with WIDTH and the corresponding synthetic spectra
computed with
SYNTHE. Examination of the results showed that
this was due to the fact that
SYNTHE read as input a model in which the elemental abundances
had been changed according to the results of WIDTH, whereas
WIDTH read a copy of the model in which all abundances were
solar scaled.
This is common practice in abundance analysis using ODF-based
model atmospheres: one computes a model-atmosphere with a given
ODF and the corresponding set of abundances, then with this model
atmosphere one computes the detailed line transfer
(with SYNTHE, WIDTH or similar
codes), varying the elemental abundances until a match is
found between computed and observed spectra/EWs. 

This practice
is legitimate inasmuch as the abundance variation from
the set assumed in the model computation has little or no
influence on the model structure. While true for trace
elements, it is in general not true for $\alpha$ elements,
as pointed out by \citet{BC03}. Indeed, both the WIDTH and
the SYNTHE suite, upon reading the input model atmosphere and
abundances, compute the number densities of atomic species
at all depths. Thus while a change in Fe abundance will have
a negligible effect on the model structure, a change in Mg abundance
will result in a change in the model ionization structure, since Mg
is the main electron donor. It must be stressed that in such cases the
results obtained by computing synthetic spectra from a model computed
with a {\em different} chemical composition are inconsistent.
On the other hand, in those cases in which e.g. a computation of abundance
with WIDTH, yields the same result both when solar scaled abundances
are used as input (and thus in the computation of number densities)
and when the abundances so derived are used as input, it is
justified to assume that the abundance of the given element has no significant
effect on the model structure.

This assumption is indeed adequate for all
the stars in our sample {\em except} Star \# \object{1665}.
For this star we computed
an ad-hoc opacity sampling model with observed abundances by using
ATLAS 12 (\citealt{kurucz93}).
In Fig. \ref{models} we show the temperature and ionization structure
for models appropriate for Star \# \object{1665} in three cases:
1) ATLAS 12 opacity sampling model with abundances given in Table
\ref{abund_abs}; 2) ATLAS 9 ODF-based model with solar scaled abundances;
3) ATLAS 9 ODF-based model  with abundances given in Table
\ref{abund_abs}.
It is obvious that changing the abundances in the ATLAS 9 model
alters the ionization structure, which is, however, different from
the one consistently computed with ATLAS 12 when all the final
abundances are taken into account during the model computation.
From Fig. \ref{models} one may appreciate that 
using the different models would imply a difference
in derived Mg abundance on the order of 0.2 dex.
The process is clearly iterative, as one starts from abundances
derived from a model with solar-scaled abundances, then uses these to compute
an ATLAS 12 model and finally determines new abundances, to be used for a new
ATLAS 12 model; and so on, until the derived
abundances coincide with the abundances used in the
model computation. In our case two  iterations were sufficient.
We want to stress that the need for computing
an ATLAS 12 model for this star is due to
its extremely low gravity;
nevertheless, the derived abundances appear in reassuring concordance
with the ones derived for the other stars.

The atomic data used in the analysis are provided in Table
\ref{abund1}; they are almost always the same as in Paper I (derived from the Castelli -- Kurucz database, \citealt{castelli03})
except for \ion{Mg}{i}, for which we   adopted the log gf values derived by
\citet{gratton03} from measured lifetimes and theoretical
branching ratios, and \ion{Ca}{i}, for which  we adopted the
furnace measurements of \citet{SR}.
All the lines have either experimental or theoretical
$\log(gf)$ values, with the exception of four
\ion{Si}{i} lines (marked as ED in Table \ref{abund1}) which have
oscillator strengths derived from solar spectra (\citealt{edvardsson93}).


\section{Results}

The derived abundances are listed in Table \ref{abund_abs} and the
corresponding abundance ratios in Table \ref{abund_rat}. The associated errors
are the scattering between the abundance given by the single lines and their
propagation in the abundance ratios. 
Iron abundances appear to be in very good agreement with the results in
\citet{tarantella04}, leading to a mean value of [Fe/H]$\sim -0.6$.
Our analysis provides [Mg/Fe]$ = -0.12$ for star
\# 1665 and  $-0.10$ for \# 1708,
coherent with the values
we found for the first three stars.
However this is significantly at odds
with the \citet{tarantella04} results.
Leaving aside the case of Star \# 1665  due to
its very low gravity and temperature, mean Mg and Ti abundances in
\citet{tarantella04} are about 0.25 dex higher than ours, while their Si
abundance is 0.17 dex {\em lower} than the one we obtained.

The three stars in common testify that no significant offset is introduced by
the different T$_{eff}$ calibration methods. The (almost) identical gravities
derived starting with (almost) identical iron abundances lead us to infer that
little difference should exist between the structure of our ATLAS model and of
the MARCS ones used in \citet{tarantella04}.
A comparison of EWs for lines in common suggests that there
is no significant offset in EW measurements either: a linear regression provides slope of
about 1 and no significant offset with an RMS,
which is on the order of our measurement error (see Fig. \ref{pl_ew}). The lines in common between the two samples were
1 for \ion{Mg}{i}, 3 for \ion{Si}{i}, 4 for \ion{Ti}{i}, 2 for \ion{Ca}{i}, 7 for \ion{Fe}{i}, 8 for \ion{Fe}{ii}, and 11 for \ion{Ni}{i}.
Therefore probably the cause of this discrepancy resides in the choice
of lines used and the atomic data.


\begin{table*}

\caption{Abundances: S/N are per pixel. On Star \# \object{1282} the values are DIC1 and DIC2, respectively}
\label{abund_abs}
\begin{center}
{\scriptsize
\begin{tabular}{lllclclclclclclc}
\hline
\\
Star     & S/N & A(\ion{Fe}{i}) & $n $& A(\ion{Fe}{ii}) & $n$ & A(Mg) & $n$ & A(Si)& $n$ & A(Ca) & $n$& A(Ti) & $n$ & A(Ni) & $n$\\
         & @650nm &     &     &         &     &       &     &      &     &       &    &       &     &       &    \\
\hline
\\
Sun    &       & 7.50          &    & 7.50             &    & 7.58            &   & 7.55           &   &  6.36         &   &  6.36         &   &  6.36          \\
\object{1272}   & 43    & $6.83\pm0.08$ & 12 & $6.85\pm0.06$    & 4  & $6.88\pm0.13$   & 4 & $6.91\pm0.16$  & 4 & $5.88\pm0.09$ & 8 & $4.47\pm0.09$ & 5 & $5.32\pm0.09$ & 5 \\
\object{1282}   & 55/65 & $6.96\pm0.11$ & 41 & $6.92\pm0.10$    & 12 & $ 6.88\pm 0.05$ & 5 & $7.06\pm 0.11$ & 8 & $5.74\pm0.09$ & 8 & $4.50\pm0.13$ &16 & $5.47\pm0.08$ &12 \\
\object{1515}   & 45    & $6.94\pm0.09$ & 11 & $6.89\pm0.08$    & 4  & $6.99\pm0.09$   & 3 & $7.14\pm0.05$  & 5 & $5.85\pm0.10$ &  9& $4.52\pm0.07$ & 6 & $5.54\pm0.05$ & 6 \\
\object{1665}   & 60    & $6.99\pm0.14$ & 33 & $6.98\pm0.09$    & 8  & $6.93\pm0.12$   & 3 & $7.00\pm0.12$  & 8 & $5.79\pm0.11$ & 5 & $4.59\pm0.21$ & 11& $5.49\pm0.13$ & 8  \\
\object{1708}   & 65    & $6.94\pm0.12$ & 34 & $6.96\pm0.06$    & 5  & $6.84\pm0.08$   & 3 & $6.98\pm0.20$  & 6 & $5.65\pm0.05$ & 5 & $4.39\pm0.11$ & 16& $5.52\pm0.11$ & 15\\
\\
\hline
\\
\end{tabular}
}
\end{center}
\end{table*}

\begin{table*}

\caption{Errors in the abundances of Star \# \object{1282} due to uncertainties
in the atmospheric parameters}
\label{errors}
\begin{center}
\begin{tabular}{lrrrrrrr}
\hline
\\
                                  & $\Delta$A(\ion{Fe}{i})   & $\Delta$A(\ion{Fe}{i}I)  &$\Delta$A(Mg)     & $\Delta$A(Si)    & $\Delta$A(Ca)    & $\Delta$A(Ti)    & $\Delta$A(Ni)    \\
\hline
\\
$\Delta \xi =  \pm 0.2$ kms$^{-1}$ &$^{-0.07}_{+0.08}$&$^{-0.05}_{+0.04}$&$^{-0.03}_{-0.04}$&$\mp 0.03$        &$^{-0.12}_{+0.13}$&$^{-0.09}_{+0.10}$&$\mp 0.07$        \\
\\
$\rm \Delta T_{eff} = \pm 100$ K  &$^{+0.01}_{+0.00}$&$^{-0.15}_{+0.16}$&$^{+0.03}_{+0.00}$&$^{-0.08}_{+0.11}$&$^{+0.11}_{-0.12}$&$^{+0.18}_{-0.17}$&$^{+0.00}_{-0.01}$\\
\\
$\rm \Delta T_{eff} = \pm 100~K~~\Delta \log g = \pm 0.20$&$^{+0.07}_{-0.04}$&$^{+0.03}_{+0.04}$&$^{+0.04}_{-0.02}$&$^{+0.00}_{+0.04}$&$\pm 0.11        $&$\pm 0.019       $&$^{+0.08}_{-0.05}$\\
\\
$\Delta \log g = \pm 0.50 $       &$^{+0.09}_{-0.10}$&$^{+0.24}_{-0.30}$&$^{+0.04}_{-0.03}$&$^{+0.12}_{-0.14}$&$^{-0.01}_{+0.00}$&$^{+0.02}_{-0.03}$&$^{+0.11}_{-0.13}$\\
\\
\hline
\\
\end{tabular}
\end{center}
\end{table*}

\begin{table*}
\caption{Abundance ratios. The mean cluster values are obtained by weighted mean.}
\label{abund_rat}
\begin{center}
\begin{tabular}{lrrrrrr}
\hline
\\
Star        & \multispan1{\hfill[Fe/H]\hfill}& [Mg/Fe] & [Si/Fe] & [Ca/Fe] & [Ti/Fe] & [Ni/Fe] \\
\\
\hline
\\
\object{1272}  &    $- 0.67 \pm 0.08$ & $-0.03 \pm 0.15$ & $ 0.03 \pm 0.18$ & $ 0.19 \pm 0.12$ &  $ 0.12 \pm 0.11$ & $-0.26 \pm 0.12$  \\
\object{1282}  &    $- 0.54 \pm 0.11$ & $-0.16 \pm 0.12$ & $ 0.05 \pm 0.16$ & $-0.08 \pm 0.14$ &  $ 0.02 \pm 0.17$ & $-0.24 \pm 0.14$  \\
\object{1515}  &    $- 0.56 \pm 0.09$ & $-0.03 \pm 0.13$ & $ 0.15 \pm 0.10$ & $ 0.05 \pm 0.13$ &  $ 0.06 \pm 0.11$ & $-0.15 \pm 0.10$  \\
\object{1665}  &    $- 0.51 \pm 0.14$ & $-0.14 \pm 0.18$ & $ -0.04\pm 0.18$ & $-0.06 \pm 0.18$ &  $  0.08\pm 0.25$ & $-0.25\pm 0.19$  \\
\object{1708}  &    $- 0.56 \pm 0.12$ & $-0.18 \pm 0.14$ & $-0.01 \pm 0.23$ & $-0.15 \pm 0.13$ &  $-0.07 \pm 0.16$ & $-0.17 \pm 0.16$  \\
{\bf Mean} & {\boldmath $-0.59 \pm 0.07$} & {\boldmath $-0.11 \pm 0.07$} & {\boldmath $ 0.07 \pm 0.09$} & {\boldmath $-0.00 \pm 0.14$} & {\boldmath $-0.05 \pm 0.07$} &{\boldmath $-0.19 \pm 0.05$} \\
\\
\hline
\end{tabular}
\\
\end{center}
\end{table*}

\section{Implications for \object{Terzan 7} origin and evolution}


\subsection{{$\alpha$} elements: Mg, Si, Ca, Ti}

These elements show {\em no enhancement} with respect to
iron. 
   \begin{figure}
   \centering
   \includegraphics[width=9cm]{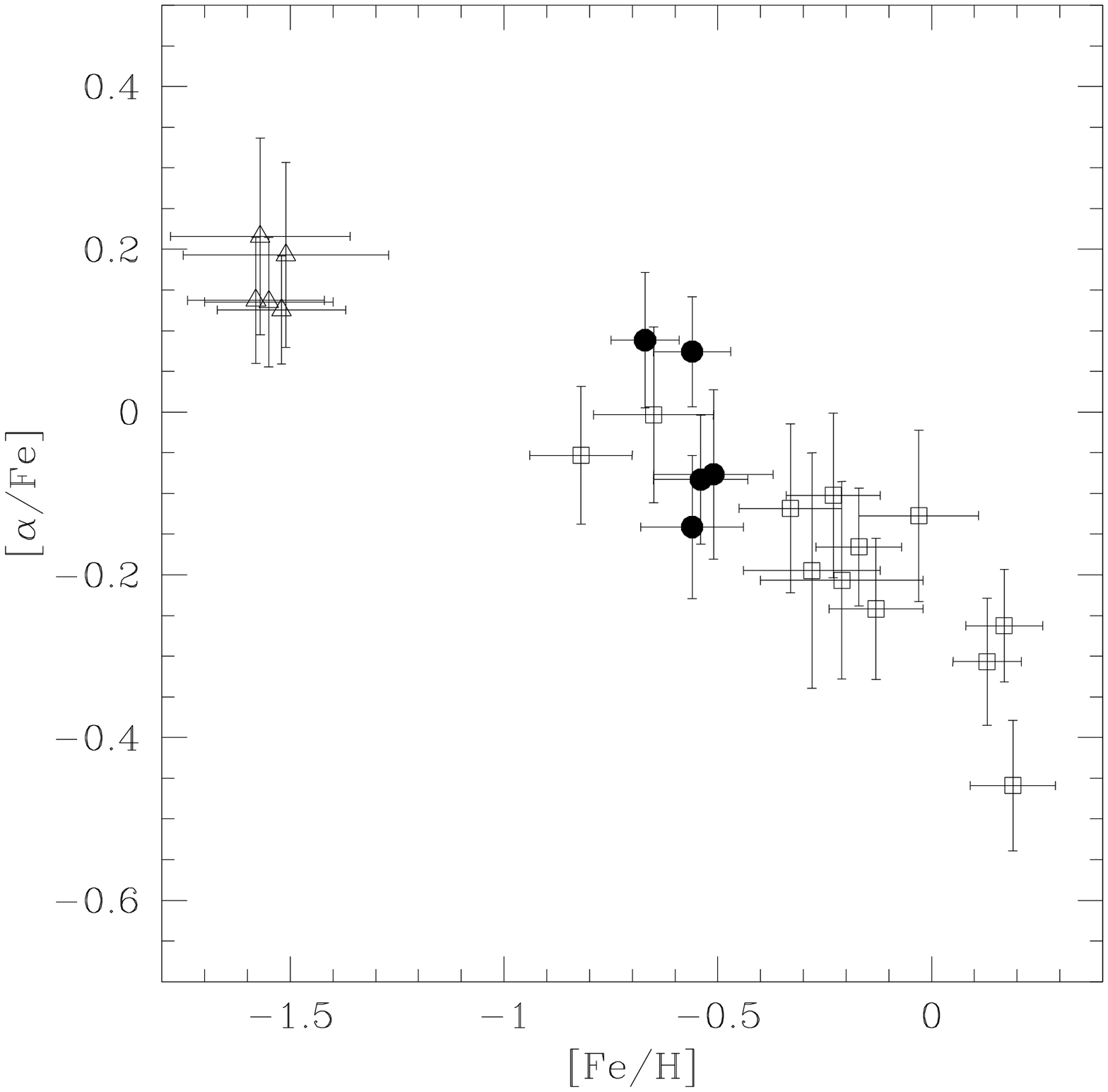}
      \caption{[$\alpha$/Fe] vs. [Fe/H]
for the five stars of \object{Terzan 7} (filled dots),
the 12 stars in the \object{Sgr dSph} main
      body studied in Paper I 
(open squares), and the five stars in \object{M54} from \citet{brown99} (open triangles). $\alpha$ is defined as the mean
      of Mg, Si, and Ca.
              }
         \label{alphasufe}
   \end{figure}
 From Fig. \ref{alphasufe} one may also see
that there is a hint of dispersion in the
abundances of $\alpha$ elements
within the cluster. Such a dispersion would be unusual in such a small globular cluster,
so that given the size of the errors and the small number of analyzed stars, for the time being we are
reluctant to claim it as real. A larger sample would be necessary to clarify the situation.

The tendency of dwarf galaxies to show low alpha content (at any given
metallicity) compared to the MW is recently emerging as a distinctive
feature of their chemical composition (\citealt{tolstoy03,venn04}).
As shown in Fig. \ref{alphasufe}, \object{Terzan 7}
stars appear to follow perfectly the same [$\alpha$/Fe] vs. [Fe/H] path
as the \object{Sgr dSph} body and \object{M 54}, which is completely included in the 
nucleus of the \object{Sgr dSph}: see \citet{Monaco04}. 
As shown in Paper I, the same pattern is also followed by the other LG dwarf
galaxies, which is strongly hints at common origins for \object{Terzan 7} and the 
Sgr dSph main body in an environment with low star formation
rate along a prolonged time.
It is also worth noticing that strongly
similar results have been reported for \object{Pal 12} (\citealt{cohen03}), which 
is
supposed to have 
formed inside the \object{Sgr dSph} and
consequently been stripped by the MW (\citealt{bellazzini03}). Indeed, the strong
chemical similarities between \object{Pal 12} and the \object{Sgr dSph} are considered by Cohen et
al. as strongly enforcing this theory of the origin of the cluster.

Similar chemical characteristics are also shared by \object{Ruprecht 106}
(\citealt{brown97}) and \object{Pal 5} (\citealt{pal5}). These clusters do not have orbital
characteristics compatible with a recent stripping  from the \object{Sgr dSph}, although
\object{Pal 5} orbit may be compatible with an older stripping event; see for this
cluster \citet{odenkirchen}. Nevertheless, they are generally considered to
have been accreted by MW from some merging event.

\subsection{Nickel}
 As an ``iron peak'' element, Ni is usually considered as enriching
the interstellar medium through the same processes as iron does and, as a
consequence, showing a stable ratio with Fe. This is typically true in Milky Way
field stars and in the large majority of galactic globular clusters
(\citealt{cayrel03}, \citealt{sneden04} and references therein). Some significant
exceptions exist and have been  known for some time, the most notable being again
\object{Pal 12} (\citealt{cohen03}) and \object{Ru 106} (\citealt{sneden04,brown97}).

Measurements are now also known for \object{Terzan 7} (this paper and
\citealt{tarantella04}) 
and the \object{Sgr dSph} body (\citealt{sbordone04_2}, \citealt{sbordone04d}, and
\citealt{bonifacio00}). In {\em all these objects} Ni appears to be
systematically underabundant with respect to iron of amounts between 0.1 to 
0.3 dex. Even if slight, this underabundance appears to significantly exceed
the spread encountered inside the Milky Way.
We have no hypothesis for the possible cause
for this underabundance,
but, again, we find here a very
distinctive signature  that ties \object{Ter 7} (and \object{Pal 12}, of course) to the \object{Sgr dSph}.

\section{Discussion} We believe that convincing evidence exists  that
the chemical composition  of the \object{Sgr dSph}
correlates strongly with that of  its present or past
(supposed) satellites. For the objects presently not dynamically linked to the
dwarf galaxy (\object{Pal 12} and maybe \object{Pal 5}) this chemical affinity can also be
considered as strongly supporting the actual existence of a past connection.
Intriguingly \object{Ru 106} is not supposed to be associated to the \object{Sgr dSph},
nevertheless it shows a very similar composition pattern. Although an
``extreme'' object among the LG dwarf galaxies due to its high metallicity, the
\object{Sgr dSph} seems to share with them a common evolutionary path. If \object{Ru 106} were the
residual of another merging process, this similarity would enforce the
interpretation that the chemical composition of
\object{Sgr dSph} is actually not peculiar at all.
The problem remains to explain {\em how} this abundance
pattern actually arises. The
\object{Sgr dSph} system interacts strongly with the MW,
and its mass, likely, decreased significantly in the process. Its
ability to retain the SN yields (and more generally the star-forming gas) is
consequently difficult to state without detailed simulations. The timescale of
their release (faster for Type II SNe, \citealt{tinsley,sw92,chiaki}) 
may also have played a role.
A final interesting aspect is the formation time of the various globular
clusters. \object{M 54}, \object{Pal 5}, \object{Pal 12}, and \object{Ter 7} lie (more or less) equally spaced along the
[$\alpha$/Fe] vs. [Fe/H] graph, leading one to think that they should have formed one after the other throughout a considerable time span, as long as [Fe/H]
may be considered to be a proxy of time.
How could a small object like \object{Sgr dSph} continue to
form globular clusters while evolving along almost 1 dex in [Fe/H] ? The strong
interaction with the MW may be an explanation, that might be telling us
something about the mechanisms of GC formation during interactions.

\begin{acknowledgements}
We are grateful to Fiorella Castelli for many interesting
discussions on model-atmospheres and for guiding us
in the use of ATLAS12.
This research was done with support from the
Italian MIUR COFIN2002 grant
``Stellar populations in the Local Group
as a tool to understand galaxy formation and evolution'' (P.I. M. Tosi).

\end{acknowledgements}


\begin{table*}
\begin{center}
\caption{Lines}
\label{abund1}
{\linespread{1.}
{\scriptsize
\begin{tabular}{rrrlrrrrrrrrrr}
\hline
\\
Ion  & $\lambda$ & log gf  & source of   & EW      & $\epsilon$ & EW       & $\epsilon$ & EW      & $\epsilon$ & EW      & $\epsilon$ & EW      & $\epsilon$\\
     & (nm)      &         & log gf      & (pm)$^a$&            & (pm)$^a$ &            & (pm)$^a$&            & (pm)$^a$&             &(pm)$^a$&           \\
     &           &         & (see notes) & \object{1272}    &            & \object{1282}     &            & \object{1515}    &            & \object{1665}    &             & \object{1708}   &           \\
\\
\ion{Fe}{i}& 466.1535 &  -1.270 & FMW &   --   &  --    &   4.21 &  6.68  &   --   &   --    &   6.53 &   7.07  &   --   &   --   \\
\ion{Fe}{i}& 479.4354 &  -4.050 & FMW &   --   &  --    &   5.90 &  6.94  &   --   &   --    &   6.25 &   6.77  &   --   &   --   \\
\ion{Fe}{i}& 484.9668 &  -2.680 & FMW &   --   &  --    &   --   &  --    &   --   &   --    &   4.50 &   6.70  &   --   &   --   \\
\ion{Fe}{i}& 489.2859 &  -1.290 & FMW &   7.11 &  6.923 &   7.15 &  6.82  &   6.51 &   6.90  &   --   &   --    &   --   &   --   \\
\ion{Fe}{i}& 489.6439 &  -2.050 & FMW &   --   &  --    &   7.20 &  7.15  &   --   &   --    &   7.80 &   7.16  &   --   &   --   \\
\ion{Fe}{i}& 491.8013 &  -1.360 & FMW &   --   &  --    &   6.41 &  6.75  &   --   &   --    &   6.77 &   6.73  &   --   &   --   \\
\ion{Fe}{i}& 510.9652 &  -0.980 & FMW &   8.24 &  6.94  &   --   &  --    &   8.24 &   7.03  &   --   &   --    &   --   &   --   \\
\ion{Fe}{i}& 552.5544 &  -1.330 & FMW &   7.32 &  6.95  &   8.05 &  6.99  &   6.66 &   6.93  &   --   &   --    &   --   &   --   \\
\ion{Fe}{i}& 585.6087 &  -1.640 & FMW &   4.11 &  6.71  &   5.84 &  6.96  &   4.20 &   6.84  &   --   &   --    &   --   &   --   \\
\ion{Fe}{i}& 585.8779 &  -2.260 & FMW &   2.26 &  6.84  &   3.55 &  7.05  &   1.95 &   6.86  &   --   &   --    &   --   &   --   \\
\ion{Fe}{i}& 586.1109 &  -2.450 & FMW &   1.44 &  6.87  &   --   &  --    &   --   &   --    &   --   &   --    &   --   &   --   \\
\ion{Fe}{i}& 587.7794 &  -2.230 & FMW &   --   &  --    &   --   &  --    &   3.24 &   7.09  &   --   &   --    &   --   &   --   \\
\ion{Fe}{i}& 588.3817 &  -1.360 & FMW &   7.83 &  6.71  &   9.72 &  6.98  &   7.89 &   6.85  &   --   &   --    &   --   &   --   \\
\ion{Fe}{i}& 595.2717 &  -1.440 & FMW &   --   &  --    &   9.04 &  6.95  &   --   &   --    &   8.37 &   6.70  &   9.59 &   6.98  \\
\ion{Fe}{i}& 601.2210 &  -4.200 & FMW &   --   &  --    &   8.38 &  7.14  &   --   &   --    &   --   &   --    &   7.83 &   6.99  \\
\ion{Fe}{i}& 601.5245 &  -4.680 & FMW &   --   &  --    &   --   &  --    &   --   &   --    &   --   &   --    &   3.44 &   6.76  \\
\ion{Fe}{i}& 601.9366 &  -3.360 & FMW &   --   &  --    &   --   &  --    &   --   &   --    &   3.09 &   7.05  &   1.93 &   6.92  \\
\ion{Fe}{i}& 602.4058 &  -0.120 & FMW &   --   &  --    &  11.97 &  6.95  &   --   &   --    &  11.65 &   6.82  &  11.97 &   6.86  \\
\ion{Fe}{i}& 602.7051 &  -1.210 & FMW &   --   &  --    &   9.66 &  6.96  &   --   &   --    &  10.18 &   6.95  &   9.66 &   6.87  \\
\ion{Fe}{i}& 605.6005 &  -0.460 & FMW &   --   &  --    &   7.92 &  6.74  &   --   &   --    &   8.29 &   6.76  &   9.24 &   6.92  \\
\ion{Fe}{i}& 607.9008 &  -1.120 & FMW &   --   &  --    &   6.07 &  6.94  &   --   &   --    &   --   &   --    &   7.02 &   7.06  \\
\ion{Fe}{i}& 609.6663 &  -1.930 & FMW &   --   &  --    &   6.65 &  6.96  &   --   &   --    &   7.28 &   6.97  &   6.79 &   6.94  \\
\ion{Fe}{i}& 610.5128 &  -2.050 & FMW &   --   &  --    &   2.28 &  6.99  &   --   &   --    &   3.19 &   7.11  &   2.38 &   6.99  \\
\ion{Fe}{i}& 612.0246 &  -5.950 & FMW &   --   &  --    &   --   &  --    &   --   &   --    &   --   &   --    &   6.37 &   6.67  \\
\ion{Fe}{i}& 615.1616 &  -3.299 & FMW &  10.45 &  6.75  &   --   &  --    &   9.99 &   6.85  &   --   &   --    &   --   &   --    \\
\ion{Fe}{i}& 615.9375 &  -1.970 & FMW &   --   &  --    &   --   &  --    &   --   &   --    &   --   &   --    &   2.93 &   7.11  \\
\ion{Fe}{i}& 616.5360 &  -1.550 & FMW &   6.11 &  6.78  &   7.06 &  6.87  &   6.24 &   6.92  &   8.08 &   6.96  &   7.26 &   6.85  \\
\ion{Fe}{i}& 618.7989 &  -1.720 & FMW &   6.77 &  6.81  &   8.20 &  6.98  &   --   &   --    &   8.64 &   6.95  &   7.72 &   6.83  \\
\ion{Fe}{i}& 622.6734 &  -2.220 & FMW &   --   &  --    &   6.21 &  7.03  &   --   &   --    &   7.19 &   7.09  &   6.31 &   7.00  \\
\ion{Fe}{i}& 649.6465 &  -0.570 & FMW &   7.25 &  6.836 &   6.99 &  6.73  &   7.39 &   6.97  &   --   &   --    &   7.49 &   6.76  \\
\ion{Fe}{i}& 651.8366 &  -2.750 & FMW &   --   &  --    &  11.40 &  7.02  &   --   &   --    &  13.03 &   7.10  &  10.42 &   6.77  \\
\ion{Fe}{i}& 659.7559 &  -1.070 & FMW &   --   &  --    &   5.86 &  7.02  &   --   &   --    &   6.54 &   7.10  &   5.99 &   7.00  \\
\ion{Fe}{i}& 670.3565 &  -3.160 & FMW &   7.81 &  6.83  &   9.42 &  6.95  &   8.20 &   7.07  &   --   &   --    &   --   &   --    \\
\ion{Fe}{i}& 672.5356 &  -2.300 & FMW &   --   &  --    &   4.39 &  7.06  &   --   &   --    &   5.29 &   7.11  &   --   &   --    \\
\ion{Fe}{i}& 673.9521 &  -4.950 & FMW &   --   &  --    &   8.63 &  6.90  &   --   &   --    &  10.55 &   6.91  &   --   &   --    \\
\ion{Fe}{i}& 674.6954 &  -4.350 & FMW &   --   &  --    &   3.42 &  6.92  &   --   &   --    &   4.84 &   6.94  &   3.52 &   6.92  \\
\ion{Fe}{i}& 679.3258 &  -2.470 & FMW &   --   &  --    &   3.78 &  7.08  &   --   &   --    &   3.97 &   7.01  &   2.77 &   6.86  \\
\ion{Fe}{i}& 684.2685 &  -1.320 & FMW &   --   &  --    &   6.36 &  7.14  &   --   &   --    &   6.71 &   7.14  &   6.12 &   7.05  \\
\ion{Fe}{i}& 684.3655 &  -0.930 & FMW &   --   &  --    &   8.33 &  6.98  &   --   &   --    &   8.74 &   6.99  &   8.40 &   6.93  \\
\ion{Fe}{i}& 685.7250 &  -2.150 & FMW &   --   &  --    &   --   &  --    &   --   &   --    &   --   &   --    &   4.10 &   6.79  \\
\ion{Fe}{i}& 686.2492 &  -1.570 & FMW &   --   &  --    &   5.15 &  7.06  &   --   &   --    &   5.92 &   7.13  &   4.90 &   6.98  \\
\ion{Fe}{i}& 691.6680 &  -1.450 & FMW &   --   &  --    &   8.68 &  7.02  &   --   &   --    &  10.25 &   7.20  &   8.96 &   7.00  \\
\ion{Fe}{i}& 697.6922 &  -1.850 & FMW &   --   &  --    &   2.44 &  6.85  &   --   &   --    &   3.71 &   7.04  &   2.72 &   6.89  \\
\ion{Fe}{i}& 698.8524 &  -3.660 & FMW &   --   &  --    &   --   &  --    &   --   &   --    &  12.04 &   7.12  &  10.40 &   7.02  \\
\ion{Fe}{i}& 702.2952 &  -1.250 & FMW &   --   &  --    &   9.64 &  7.04  &   --   &   --    &  10.62 &   7.11  &   8.96 &   6.84  \\
\ion{Fe}{i}& 706.9531 &  -4.340 & FMW &   --   &  --    &   4.00 &  6.93  &   --   &   --    &   5.03 &   6.87  &   --   &   --    \\
\ion{Fe}{i}& 721.9682 &  -1.690 & FMW &   --   &  --    &   8.16 &  7.05  &   --   &   --    &   9.27 &   7.14  &   8.74 &   7.08  \\
\ion{Fe}{i}& 747.6375 &  -1.680 & FMW &   --   &  --    &   2.71 &  7.00  &   --   &   --    &   --   &   --    &   3.73 &   7.17  \\
\ion{Fe}{i}& 749.8530 &  -2.250 & FMW &   --   &  --    &   3.87 &  6.94  &   --   &   --    &   5.50 &   7.12  &   4.23 &   6.97  \\
\ion{Fe}{i}& 758.3787 &  -1.990 & FMW &   --   &  --    &  14.44 &  6.90  &   --   &   --    &  15.74 &   6.90  &  15.71 &   7.00  \\
\ion{Fe}{i}& 794.1087 &  -2.580 & FMW &   --   &  --    &   9.04 &  6.91  &   --   &   --    &  10.19 &   6.92  &   9.24 &   6.87  \\
\ion{Fe}{i}& 883.8427 &  -1.980 & FMW &   --   &  --    &  17.12 &  6.92  &   --   &   --    &   --   &   --    &  18.18 &   6.97  \\
\\
\ion{Fe}{ii}& 483.3197 &  -4.780 & FMW &   2.22 &  6.84  &   --   &  --    &   --   &   --    &     -- &     --  &     -- &     --  \\
\ion{Fe}{ii}& 499.3358 &  -3.650 & FMW &   --   &  --    &   6.27 &  7.121 &   4.65 &   6.88  &     -- &     --  &     -- &     --  \\
\ion{Fe}{ii}& 510.0664 &  -4.370 & FMW &   3.54 &  6.96  &   --   &  --    &   --   &   --    &     -- &     --  &     -- &     --  \\
\ion{Fe}{ii}& 513.2669 &  -4.180 & FMW &   --   &  --    &   3.50 &  7.013 &   3.01 &   7.01  &     -- &     --  &     -- &     --  \\
\ion{Fe}{ii}& 525.6937 &  -4.250 & K88 &   3.08 &  6.82  &   2.90 &  7.035 &   --   &   --    &     -- &     --  &     -- &     --  \\
\ion{Fe}{ii}& 526.4812 &  -3.190 & FMW &   5.71 &  6.79  &   4.55 &  6.809 &   4.56 &   6.90  &     -- &     --  &     -- &     --  \\
\ion{Fe}{ii}& 599.1376 &  -3.557 & K88 &   --   &  --    &   3.23 &  6.798 &   --   &   --    &   3.90 &   7.04  &   3.91 &   6.849 \\
\ion{Fe}{ii}& 608.4111 &  -3.808 & K88 &   --   &  --    &   2.74 &  6.977 &   --   &   --    &   2.23 &   6.87  &   --   &     --  \\
\ion{Fe}{ii}& 614.9258 &  -2.724 & K88 &   --   &  --    &   3.16 &  6.864 &   2.75 &   6.79  &   3.58 &   7.10  &   4.17 &   7.000 \\
\ion{Fe}{ii}& 624.7557 &  -2.329 & K88 &   --   &  --    &   4.50 &  6.828 &   --   &   --    &   4.32 &   6.92  &   5.80 &   6.979 \\
\ion{Fe}{ii}& 636.9462 &  -4.253 & K88 &   --   &  --    &   2.28 &  6.918 &   --   &   --    &   2.43 &   7.01  &   2.84 &   6.970 \\
\ion{Fe}{ii}& 643.2680 &  -3.500 & H92 &   --   &  --    &   4.60 &  6.760 &   --   &   --    &   4.50 &   7.02  &     -- &     --  \\
\ion{Fe}{ii}& 645.6383 &  -2.075 & K88 &   --   &  --    &   5.87 &  6.907 &   --   &   --    &   4.90 &   6.84  &     -- &     --  \\
\ion{Fe}{ii}& 651.6080 &  -3.380 & H92 &   --   &  --    &   6.08 &  6.953 &   --   &   --    &   5.65 &   7.02  &   6.81 &   7.021 \\
\hline
\\
\multispan{10}{$a)$ 1 pm = 10$^{-12}$ m 
\hfill}\\
\\
\multispan{10}{FMW   \citet{fuhr88}  \hfill}\\
\multispan{10}{H92   \citet{hannaford92}  \hfill}\\
\multispan{10}{K88   \citet{kurucz88} \hfill}\\
\end{tabular}
}}
\end{center}
\end{table*}

\addtocounter{table}{-1}

\begin{table*}
\begin{center}
\caption{Lines (continued)}
{\linespread{1.}
{\scriptsize
\begin{tabular}{rrrlrrrrrrrrrr}
\hline
\\
Ion  & $\lambda$ & log gf  & source of   & EW      & $\epsilon$ & EW       & $\epsilon$ & EW      & $\epsilon$ & EW      & $\epsilon$ & EW      & $\epsilon$\\
     & (nm)      &         & log gf      & (pm)$^a$&            & (pm)$^a$ &            & (pm)$^a$&            & (pm)$^a$&             &(pm)$^a$&           \\
     &           &         & (see notes) & \object{1272}    &            & \object{1282}     &            & \object{1515}    &            & \object{1665}    &             & \object{1708}   &           \\
\\
\hline
\\
\ion{Mg}{i}& 552.8405&  -0.522  & GC   &   20.79  &   6.88   &  22.72   &  6.86   &   --    &   --     &   --    &   --      &   --    &   --  \\
\ion{Mg}{i}& 571.1087&  -1.729  & GC   &   12.14  &   7.03   &  12.75   &  6.96   &  12.37  &   7.12   &   --    &   --      &   --    &   --  \\
\ion{Mg}{i}& 631.8716&  -1.945  & GC   &    2.89  &   6.68   &   --     &  --     &   3.81  &   6.91   &   --    &   --      &   4.49  &   6.88  \\
\ion{Mg}{i}& 631.9237&  -2.165  & GC   &    3.04  &   6.93   &   3.30   &  6.91   &   2.79  &   6.93   &   3.54  &   6.88    &   3.31  &   6.90  \\
\ion{Mg}{i}& 738.7689&  -1.020  & KP   &    --    &   --     &   --     &  --     &   --    &   --     &   4.81  &   6.82    &   --    &   --  \\
\ion{Mg}{i}& 871.7825&  -0.772  & GC   &    --    &   --     &   5.40   &  6.85   &   --    &   --     &   6.41  &   7.10    &   4.76  &   6.73  \\
\ion{Mg}{i}& 892.3568&  -1.659  & GC   &    --    &   --     &   4.28   &  6.84   &   --    &   --     &   --    &   --      &   --    &   --  \\
\\
\ion{Si}{i}& 594.8541&  -1.230  & GARZ &    --    &   --     &   7.32   &  7.04   &   8.15  &   7.23   &   --    &   --      &   --    &   --  \\
\ion{Si}{i}& 612.5021&  -1.521  & ED   &    3.49  &   7.16   &   3.05   &  7.19   &   2.76  &   7.15   &   --    &   --      &   --    &   --  \\
\ion{Si}{i}& 613.1852&  -1.140  & KP   &    --    &   --     &   --     &  --     &   --    &   --     &   3.68  &   7.02    &   --    &   --  \\
\ion{Si}{i}& 614.2482&  -1.480  & ED   &    2.37  &   6.87   &   1.97   &  6.88   &   2.70  &   7.10   &   1.97  &   6.92    &   --    &   --  \\
\ion{Si}{i}& 614.5015&  -1.430  & ED   &    2.05  &   6.73   &   --     &  --     &   2.85  &   7.08   &   2.17  &   6.93    &   --    &   --  \\
\ion{Si}{i}& 615.5134&  -0.770  & ED   &    5.99  &   6.89   &   6.30   &  7.09   &   6.43  &   7.12   &   5.60  &   7.06    &   --    &   --  \\
\ion{Si}{i}& 703.4902&  -0.880  & GARZ &    --    &   --     &   4.22   &  7.14   &   --    &   --     &   --    &   --      &   5.05  &   7.22  \\
\ion{Si}{i}& 725.0627&  -1.042  & SG   &    --    &   --     &   --     &  --     &   --    &   --     &   3.45  &   6.90    &   3.84  &   6.83  \\
\ion{Si}{i}& 727.5292&  -1.003  & SG   &    --    &   --     &   4.07   &  6.90   &   --    &   --     &   4.79  &   7.14    &   3.99  &   6.82  \\
\ion{Si}{i}& 728.9176&  -0.197  & SG   &    --    &   --     &   --     &  --     &   --    &   --     &   7.36  &   6.81    &   7.96  &   6.71  \\
\ion{Si}{i}& 793.2349&  -0.470  & GARZ &    --    &   --     &   5.33   &  7.05   &   --    &   --     &   --    &   --      &   5.67  &   7.03  \\
\ion{Si}{i}& 874.2446&  -0.630  & KP   &    --    &   --     &   6.17   &  7.20   &   --    &   --     &   5.51  &   7.21    &   6.88  &   7.23  \\
\\
\ion{Ca}{i}& 551.2980&  -0.447  & SR   &   11.00  &   5.90   &  11.72   &  5.72   &   9.85  &   5.69   &   --    &   --      &   --    &   --  \\
\ion{Ca}{i}& 585.7451&   0.240  & SR   &   14.37  &   5.76   &   --     &  --     &  14.75  &   5.83   &   --    &   --      &   --    &   --  \\
\ion{Ca}{i}& 586.7562&  -1.490  & GC   &    4.87  &   5.79   &   4.73   &  5.54   &   4.71  &   5.78   &   --    &   --      &   --    &   --  \\
\ion{Ca}{i}& 616.1296&  -1.266  & SR   &   10.22  &   5.98   &   --     &  --     &   --    &   --     &   --    &   --      &   --    &   --  \\
\ion{Ca}{i}& 616.6438&  -1.142  & SR   &    --    &   --     &  12.19   &  5.87   &  10.97  &   6.05   &  13.90  & 5.90      &   --    &   --  \\
\ion{Ca}{i}& 616.9042&  -0.797  & SR   &   12.91  &   6.05   &  13.46   &  5.76   &  11.89  &   5.88   & 15.87   & 5.93      &  13.23  & 5.68  \\
\ion{Ca}{i}& 643.9075&   0.390  & SR   &   18.79  &   5.80   &  21.41   &  5.73   &  18.69  &   5.76   &   --    &   --      &  20.87  &  5.64 \\
\ion{Ca}{i}& 645.5597&  -1.290  & SR   &    --    &   --     &  10.88   &  5.75   &   9.57  &   5.90   &  11.94  &   5.63    &  10.31  &   5.63  \\
\ion{Ca}{i}& 649.3781&  -0.109  & SR   &   16.15  &   5.87   &  17.72   &  5.72   &  15.70  &   5.84   &  19.39  & 5.71      &  17.09  & 5.57  \\
\ion{Ca}{i}& 649.9650&  -0.818  & SR   &   12.17  &   5.87   &  13.96   &  5.81   &  12.22  &   5.94   &  15.41  & 5.77      &  13.59  &  5.70 \\
\\
\ion{Ti}{i}& 488.5078&   0.358  & MFW  &    --    &   --     &   --     &  --     &  10.72  &   4.46   &   --    &   --      &   --    &   --  \\
\ion{Ti}{i}& 491.5229&  -1.019  & MFW  &    5.14  &   4.58   &   --     &  --     &   4.06  &   4.44   &   8.68  &   4.45    &   5.49  &   4.30  \\
\ion{Ti}{i}& 499.7096&  -2.118  & MFW  &   11.51  &   4.35   &  14.81   &  4.58   &  11.41  &   4.48   &   --    &   --      &   --    &   --  \\
\ion{Ti}{i}& 508.7058&  -0.780  & MFW  &    9.36  &   4.53   &   --     &  --     &   9.19  &   4.57   &   --    &   --      &   --    &   --  \\
\ion{Ti}{i}& 586.6450&  -0.840  & MFW  &   11.85  &   4.37   &  15.65   &  4.57   &  12.41  &   4.61   &   --    &   --      &   --    &   --  \\
\ion{Ti}{i}& 606.4626&  -1.944  & MFW  &    --    &   --     &   9.61   &  4.50   &   --    &   --     &  13.38  &   4.70    &   9.79  &   4.53  \\
\ion{Ti}{i}& 609.2792&  -1.379  & MFW  &    --    &   --     &   4.73   &  4.40   &   --    &   --     &   6.88  &   4.31    &   3.81  &   4.28  \\
\ion{Ti}{i}& 612.6215&  -1.425  & MFW  &    9.80  &   4.52   &  13.75   &  4.71   &   9.32  &   4.56   &   --    &   --      &   --    &   --  \\
\ion{Ti}{i}& 631.2236&  -1.552  & MFW  &    --    &   --     &   7.91   &  4.41   &   --    &   --     &  10.52  &   4.36    &   6.76  &   4.27  \\
\ion{Ti}{i}& 633.6098&  -1.743  & MFW  &    --    &   --     &   7.34   &  4.50   &   --    &   --     &  10.38  &   4.52    &   6.98  &   4.47  \\
\ion{Ti}{i}& 655.6061&  -1.074  & MFW  &    --    &   --     &  12.51   &  4.62   &   --    &   --     &  16.40  &   4.86    &  10.88  &   4.35  \\
\ion{Ti}{i}& 659.9105&  -2.085  & MFW  &    --    &   --     &   --     &  --     &   --    &   --     &  15.41  &   4.83    &   9.88  &   4.38  \\
\ion{Ti}{i}& 686.1447&  -0.740  & MFW  &    --    &   --     &   4.95   &  4.27   &   --    &   --     &   7.42  &   4.25    &   5.03  &   4.31  \\
\ion{Ti}{i}& 718.8565&  -1.760  & MFW  &    --    &   --     &   7.04   &  4.37   &   --    &   --     &   --    &   --      &   5.73  &   4.22  \\
\ion{Ti}{i}& 744.0578&  -0.700  & MFW  &    --    &   --     &   5.15   &  4.20   &   --    &   --     &   --    &   --      &   4.88  &   4.19  \\
\ion{Ti}{i}& 802.4843&  -1.140  & MFW  &    --    &   --     &   --     &  --     &   --    &   --     &   --    &   --      &   7.61  &   4.42  \\
\ion{Ti}{i}& 835.3161&  -2.677  & MFW  &    --    &   --     &   9.82   &  4.61   &   --    &   --     &  14.72  &   4.75    &   7.99  &   4.42  \\
\ion{Ti}{i}& 867.5372&  -1.669  & MFW  &    --    &   --     &  14.80   &  4.58   &   --    &   --     &   --    &   --      &  14.29  &   4.49  \\
\ion{Ti}{i}& 868.2980&  -1.941  & MFW  &    --    &   --     &  11.91   &  4.46   &   --    &   --     &  17.32  &   4.72    &  11.89  &   4.46  \\
\ion{Ti}{i}& 869.2331&  -2.295  & MFW  &    --    &   --     &  10.48   &  4.63   &   --    &   --     &   --    &   --      &  10.25  &   4.61  \\
\ion{Ti}{i}& 873.4712&  -2.384  & MFW  &    --    &   --     &   9.15   &  4.56   &   --    &   --     &  14.40  &   4.74    &   8.18  &   4.47  \\
\\
\ion{Ni}{i}& 493.5830&  -0.350  & FMW  &    --    &   --     &   7.40   &  5.49   &   --    &   --     &   --    &   --      &   7.33  &   5.40  \\
\ion{Ni}{i}& 585.7746&  -0.636  & K88  &    4.37  &   5.43   &   5.36   &  5.61   &   4.47  &   5.60   &   --    &   --      &   --    &   --  \\
\ion{Ni}{i}& 600.7306&  -3.330  & FMW  &    --    &   --     &   8.42   &  5.48   &   --    &   --     &   --    &   --      &   9.12  &   5.54  \\
\ion{Ni}{i}& 608.6276&  -0.530  & FMW  &    --    &   --     &   4.01   &  5.37   &   --    &   --     &   4.67  &   5.45    &   4.83  &   5.48  \\
\ion{Ni}{i}& 611.1065&  -0.870  & FMW  &    --    &   --     &   --     &  --     &   --    &   --     &   --    &   --      &   3.47  &   5.33  \\
\ion{Ni}{i}& 612.8963&  -3.330  & FMW  &    --    &   --     &   9.04   &  5.58   &   6.28  &   5.44   &   --    &   --      &   --    &   --  \\
\ion{Ni}{i}& 613.0130&  -0.960  & FMW  &    1.43  &   5.16   &   --     &  --     &   2.25  &   5.55   &   --    &   --      &   --    &   --  \\
\ion{Ni}{i}& 617.5360&  -0.530  & FMW  &    5.01  &   5.33   &   5.32   &  5.39   &   5.35  &   5.55   &   5.34  &   5.33    &   5.93  &   5.44  \\
\ion{Ni}{i}& 617.6807&  -0.260  & WL   &    6.42  &   5.33   &   7.04   &  5.44   &   6.77  &   5.56   &   --    &   --      &   --    &   --  \\
\ion{Ni}{i}& 617.7236&  -3.500  & FMW  &    4.99  &   5.37   &   5.66   &  5.36   &   4.96  &   5.57   &   7.39  &   5.41    &   --    &   --  \\
\ion{Ni}{i}& 618.6708&  -0.960  & FMW  &    --    &   --     &   --     &  --     &   --    &   --     &   --    &   --      &   3.30  &   5.41  \\
\ion{Ni}{i}& 620.4600&  -1.100  & WL   &    --    &   --     &   2.46   &  5.38   &   --    &   --     &   2.84  &   5.39    &   2.83  &   5.43  \\
\ion{Ni}{i}& 632.7592&  -3.150  & FMW  &    --    &   --     &   --     &  --     &   --    &   --     &   --    &   --      &  11.26  &   5.69  \\
\ion{Ni}{i}& 648.2796&  -2.630  & FMW  &    --    &   --     &  10.77   &  5.51   &   --    &   --     &  12.10  &   5.51    &  10.99  &   5.46  \\
\ion{Ni}{i}& 658.6307&  -2.810  & FMW  &    --    &   --     &   --     &  --     &   --    &   --     &  12.39  &   5.76    &  11.33  &   5.72  \\
\ion{Ni}{i}& 659.8592&  -0.980  & FMW  &    --    &   --     &   --     &  --     &   --    &   --     &   --    &   --      &   2.88  &   5.50  \\
\ion{Ni}{i}& 677.2313&  -0.980  & FMW  &    --    &   --     &   6.62   &  5.47   &   --    &   --     &   7.12  &   5.47    &   7.59  &   5.57  \\
\ion{Ni}{i}& 700.1534&  -3.660  & FMW  &    --    &   --     &   --     &  --     &   --    &   --     &   --    &   --      &   5.49  &   5.55  \\
\ion{Ni}{i}& 706.2951&  -3.500  & FMW  &    --    &   --     &   5.77   &  5.50   &   --    &   --     &   7.98  &   5.61    &   6.18  &   5.52  \\
\ion{Ni}{i}& 778.8936&  -2.420  & FMW  &    --    &   --     &   --     &  --     &   --    &   --     &   --    &   --      &  14.92  &   5.71  \\
\\
\hline
\\
\multispan{10}{$a)$ 1 pm = 10$^{-12}$ m 
\hfill}\\
\\
\multispan{10}{ED    \citet{edvardsson93} \hfill}\\
\multispan{10}{FMW   \citet{fuhr88} \hfill}\\
\multispan{10}{GARZ  \citet{garz73} \hfill}\\
\multispan{10}{GC    \citet{gratton03} \hfill}\\
\multispan{10}{K88   \citet{kurucz88}  \hfill}\\
\multispan{10}{KP    \citet{kurucz75}  \hfill}\\
\multispan{10}{MFW   \citet{martin88}  \hfill}\\
\multispan{10}{NBS   \citet{wiese66}  \hfill}\\
\multispan{10}{SG    \citet{schulz69} \hfill}\\
\multispan{10}{WL    \citet{WL}   \hfill}\\
\end{tabular}
}}
\end{center}
\end{table*}

\end{document}